\pdfoutput=1
\RequirePackage{atbegshi}
\newif\ifarxivshift \arxivshifttrue
\AtBeginShipout{\ifarxivshift\global\arxivshiftfalse\setbox\AtBeginShipoutBox=\hbox{\kern6mm\box\AtBeginShipoutBox}\fi}
\documentclass[conference,a4paper]{APSIPA2026}
\usepackage{amsmath}
\usepackage{graphicx}
\usepackage{multirow}
\usepackage{array}          
\usepackage{threeparttable}
\usepackage{balance}
\usepackage[backend=biber, style=ieee,maxbibnames=5, minbibnames=1]{biblatex}
\AtEveryBibitem{\clearfield{doi}\clearfield{issn}\clearfield{isbn}\clearfield{eprintclass}}
\graphicspath{{figures/}{../figures/}}

\newcommand{\DNNAWMcut}{2.25\,kHz}
\newcommand{\DNNAWMacc}{$58.9\%$}
\newcommand{\DNNAWMpesq}{$3.34$}

\newcommand{\DNNAWMfrag}{0.62}

\usepackage{geometry}
\usepackage{fancyhdr}

\fancypagestyle{firststyle}{\fancyhf{}
	\fancyhead[C]{2026 Asia Pacific Signal and Information Processing Association Annual Summit and Conference (APSIPA ASC)}}

\usepackage[hidelinks,breaklinks=true]{hyperref}

\begin{document}
	
	\title{How Fragile Is Your Watermark?\\
		Training-Free Structural Removal of Neural Audio Watermarks}
	
	\author{
		\authorblockN{Likhith Kumara\authorrefmark{1}}
		\authorblockA{\authorrefmark{1} Indian Institute of Technology Madras, Chennai, India \\
			E-mail: da24s026@smail.iitm.ac.in}
	}
	
	\maketitle
	\thispagestyle{firststyle}
	\pagestyle{empty}
	
	{\renewcommand{\thefootnote}{}\footnotetext{Code: \url{https://github.com/i-618/audio-watermark-fragility}}}
	
	\begin{abstract}
		Neural audio watermarks are increasingly used to attribute and detect AI-generated speech, so their practical
		value rests on how cheaply an adversary can remove them. Robustness is usually measured by running a fixed
		battery of distortions blindly against every scheme. We instead make removal \emph{diagnostic}: from a few
		clean/watermarked pairs we compute cheap structural probes that reveal \emph{where} a watermark sits in the
		signal (its embedding domain), then apply a single \emph{domain-matched} attack rather than a blind sweep.
		We further summarize each scheme with one \emph{threshold-free fragility} score, the area under its
		accuracy-versus-quality trade-off, which an accuracy-only benchmark cannot provide. Across ten watermarking
		schemes the probes separate fragile from robust marks: for magnitude and carrier-domain watermarks a single
		matched attack erases the payload (WavMark, SilentCipher, \texttt{audiowmark}) or removes the detection flag
		(AudioSeal) at high objective quality (PESQ $\ge3.6$), whereas
		latent-domain marks (VoiceMark, WMCodec, AlignMark,
		AWARE) resist every training-free attack we apply. The same
		pair-only probe signatures also identify which watermarking scheme is present ($84\%$ over ten schemes).
	\end{abstract}
	
	
	\section{Introduction}
	Audio watermarking is now a front-line tool for attributing and detecting AI-generated speech, and its
	usefulness rests entirely on \emph{robustness}: a watermark protects provenance only if an adversary cannot
	quietly remove it. Modern schemes pursue robustness by embedding in very different places. The
	classical tool \texttt{audiowmark}~\cite{audiowmark} and the learned DNN-AWM~\cite{dnnawm} hide bits
	in the spectral magnitude; WavMark~\cite{wavmark}, AudioSeal~\cite{audioseal}, SilentCipher~\cite{silentcipher},
	and Timbre~\cite{timbre} place neural marks in magnitude- and vocoder-domain features; and a newer wave embeds
	in codec, latent, and speaker representations: WMCodec~\cite{wmcodec}, VoiceMark~\cite{voicemark},
	AWARE~\cite{aware}, and AlignMark~\cite{alignmark}.
	
	Robustness, however, is usually reported as a single number, bit accuracy under a fixed battery of
	distortions~\cite{audiomarkbench}, which treats every scheme the same way and collapses this diversity. It
	hides the two questions an attacker actually asks: \emph{where} in the signal does this particular mark live,
	and \emph{which} perturbation removes it most cheaply.
	
\begin{figure*}[t]\begin{center}
			\includegraphics[width=0.82\textwidth]{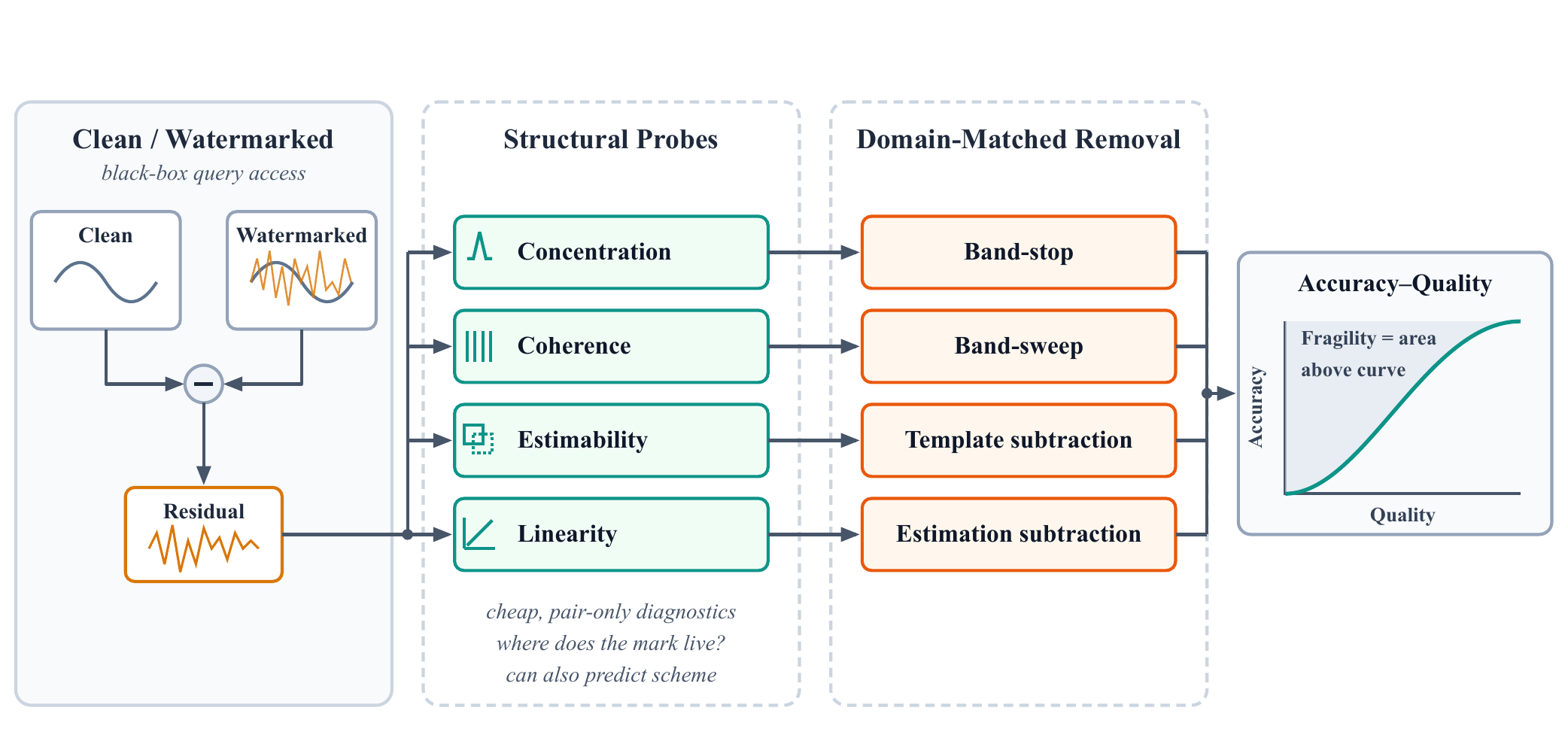}
		\end{center}
		\caption{\textbf{Method overview.} From a few clean/watermarked pairs (Sec.~\ref{sec:setup}) we
			compute the four structural probes (Sec.~\ref{sec:probes}); each probe selects a
			matched attack class (Sec.~\ref{sec:matched}), and the same probe signatures also identify which scheme
			is present. Sweeping the matched attack traces the accuracy--quality frontier whose area
			above the curve is the threshold-free fragility score (Sec.~\ref{sec:fragility}).}
		\label{fig:method}
	\end{figure*}

	We take the attacker's view and make removal \emph{diagnostic}. Different watermarks embed in different
	features: a narrowband carrier, the short-time Fourier transform (STFT) magnitude (uniform or
	frequency-redundant), a vocoder envelope, or a learned latent. Each embedding
	domain has a perturbation that is matched to it. From clean/watermarked pairs we compute cheap structural
	probes that estimate the embedding domain, and we then apply the matched attack rather than a blind battery
	(Fig.~\ref{fig:method}).
	\textbf{Contributions:}
	\begin{itemize}
		\item A \emph{diagnose-then-remove} methodology: pair-only structural probes that characterize a watermark's
		embedding domain and select a domain-matched removal attack (Sec.~\ref{sec:probes}).
		\item \emph{Demonstrated matched removals} at high objective quality (PESQ/ViSQOL) across a broad
		roster (payload erasure for WavMark, SilentCipher, and \texttt{audiowmark}, and detection-flag
		removal for AudioSeal), plus a cross-band redundancy probe that surfaces a band-sweep break for the
		otherwise subtraction-robust DNN-AWM, and latent-domain robustness anchors (VoiceMark and others) whose
		resistance the probes indicated (Sec.~\ref{sec:matched}).
		\item A supporting observation that the same probe signatures \emph{identify the watermarking scheme}
		($84\%$ over ten schemes), confirming the structure the matched attacks exploit (Sec.~\ref{sec:matched}).
	\end{itemize}
	The individual perturbations deliberately reuse
	established attack classes~\cite{voloshynovskiy,stirmark,zhao} so that any removal is attributable to the
	\emph{diagnosis} rather than to a bespoke attack.
	
	\section{Related Work}
	\textbf{Removal and robustness.} AudioMarkBench~\cite{audiomarkbench} benchmarks robustness by reporting bit
	accuracy/detection under a fixed battery of perturbations at preset strengths: a defender-side, per-model
	view. RAW-Bench~\cite{RAWBench} extends this to real-world degradations and neural
	codecs, and the SoK of~\cite{sokrobustaudiowatermarking} surveys robustness of audio watermarking under
	generative-AI pipelines; both report aggregate accuracy/detection rather than the per-scheme \emph{why}, and
	apply the same distortion set uniformly to every model. HarmonicAttack~\cite{harmonicattack} trains a
	cross-domain removal network. In contrast, we derive a \emph{matched} attack per scheme from pair-only probes
	and quantify the resulting accuracy--quality trade-off, without training. Estimation/subtraction removal is a
	classical class~\cite{voloshynovskiy}; desynchronization (time/pitch
	warping) is studied in StirMark-style audio attacks~\cite{stirmark}; for images, generative regeneration
	removes watermarks~\cite{zhao}.
	\textbf{Embedding-algorithm identification.} The nearest relative of our \emph{supporting} identification
	result is steganalysis, which infers \emph{which} algorithm produced a stego signal~\cite{steganalysis}; it is
	blind and forensic, whereas we read the isolated residual from
	clean/watermarked pairs to \emph{select a matched attack}.
	
	\section{Threat Model and Setup}\label{sec:setup}
	\textbf{Threat model.} We assume \emph{query access} to a black-box watermarking service: the adversary can
	submit chosen audio and receive its watermarked version, but does not know the internal scheme. From a few
	such (clean, watermarked) pairs the adversary computes the probes below. This is the realistic setting for a
	deployed watermarking API and is the access our pair-only probes require; it is \emph{not} blind identification
	of a single intercepted clip (which our probes cannot do, since they need the residual).
	For \emph{open-weight} schemes the encoder ships publicly, so pairs are free and unlimited and chosen-input
	access is not an assumption at all; for a \emph{closed API} it reduces to a metered budget, and that budget is
	small (below).
	
	\textbf{Query budget.} Every probe is a fixed number of encoder calls per clip:
	$1$ (Concentration), $M{=}14$ (Coherence), $A{=}3$ (Estimability) and $C{\times}(1{+}6){=}21$
	(Linearity), so $39$ in total. Fitting an attack is one-off per scheme: $15$ calls for either
	subtraction variant ($3$ clips $\times\,5$ content types), $150$ for the estimation footprint, and
	\emph{none} for band-stop or band-sweep, whose parameters follow from the probe alone. Attacking a
	further clip costs \emph{no} additional queries -- the fitted template, filter or footprint is
	reused and the target clip is given, not queried. The decoder is never called by the attack; we
	query it only to score bit accuracy. The diagnosis also survives a far smaller budget than we
	report with: reducing $M$ from $14$ to $4$ and using one clip per content type leaves the
	probe-to-attack assignment unchanged, i.e. $145$ encoder calls rather than $3{,}900$.
	
	\textbf{Watermarks.} We use a unified embed/decode interface over ten schemes at their native sample rates:
	AudioSeal, SilentCipher, Timbre, VoiceMark, AlignMark, WMCodec, DNN-AWM, WavMark, AWARE and the classical tool \texttt{audiowmark};
	\textbf{Data.} $20$ test clips each from five content
	types: LibriSpeech~\cite{librispeech} (read speech),
	VCTK~\cite{vctk} (multi-speaker), MUSDB18~\cite{musdb18} (music), VocalSet~\cite{vocalset} (singing), and a pooled synthetic-speech set
	(XTTS-v2~\cite{xtts} / F5-TTS~\cite{f5tts} / CosyVoice\,2~\cite{cosyvoice2}); $4$\,s, mono, peak-normalized; a few held-out clips per model fit the linear
	footprint basis, giving $N{=}100$ test clips per scheme ($20$ per content type $\times$ five content types), about
	$1{,}000$ watermarked clips across the ten schemes. \textbf{Quality.} We measure quality with Perceptual Evaluation of Speech Quality
	(PESQ, wideband); ``transparent''
	$\approx$ PESQ$\,>3.5$. Because PESQ is speech-tuned and unreliable for desynchronized audio
	(Sec.~\ref{sec:limits}), we lead with energy/magnitude attacks (PESQ valid) and validate the saved removal
	pairs with audio-mode ViSQOL (Virtual Speech Quality Objective Listener).
	
	\section{Probes}\label{sec:probes}
\begin{figure}[t]\begin{center}
			\includegraphics[width=\columnwidth]{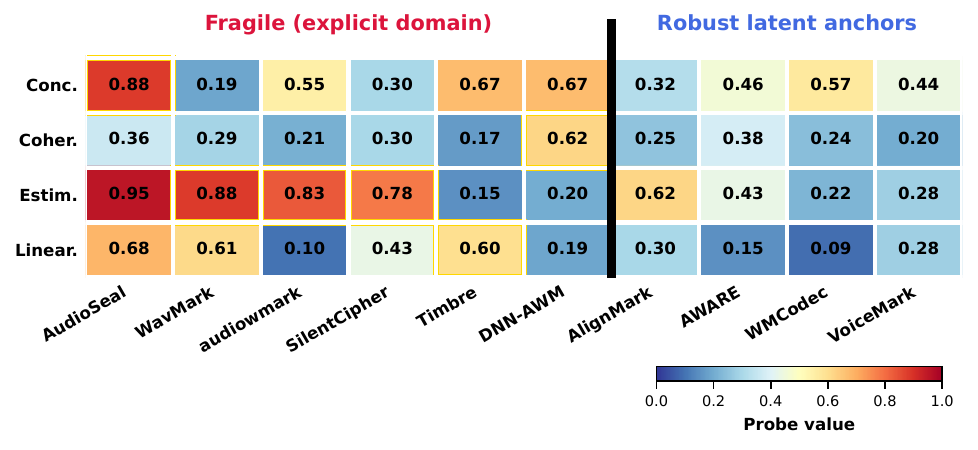}
		\end{center}
		\caption{\textbf{Structural probe signatures per scheme} (pair-only, no attack run; means over $N{=}100$
			clips/scheme, bootstrap $95\%$ CI half-widths $\le0.07$; formulas in Sec.~\ref{sec:probes}).
			The cells with relative higher value is each scheme's
			\emph{defining} probe, which selects its matched attack (Sec.~\ref{sec:matched}); the black line splits the
			fragile group (left) from the robust latent anchors (right), whose uniformly low values predict their
			resistance. The distinct signatures also identify which scheme is present (Sec.~\ref{sec:matched}).}
		\label{fig:signatures}
	\end{figure}

	All probes are computed from clean/watermarked pairs with no attack search. Each estimates \emph{where} a mark
	sits and therefore which attack class is matched to it (Fig.~\ref{fig:method}).
	Each probe is named by the structural quantity it \emph{measures}, so the
	diagnosis is non-circular: none runs the attack it selects.
	Write $r{=}y{-}h$ for the residual of a clean/watermarked pair $(h,y)$, $\hat r$ for its discrete Fourier
	transform (DFT), and $V_m$ for the
	time-averaged residual STFT magnitude under message $m$.
	\begin{itemize}
		\item \textbf{Concentration}: energy fraction of $r$ in its most concentrated band. Partitioning the $F$
		spectral bins of $P[f]{=}|\hat r[f]|^2$ into $16$ equal bands $\{B_b\}$,
		\begin{equation}
			\mathrm{Concentration}=\max_b\Big(\textstyle\sum_{f\in B_b}P[f]\Big)\Big/\textstyle\sum_f P[f].
			\label{eq:concentration}
		\end{equation}
		High $\Rightarrow$ a narrowband carrier $\Rightarrow$ \emph{band-stop} (notch) removable.
		\item \textbf{Coherence}: cross-band agreement of the payload code, whether the \emph{same} payload is
		recoverable from disjoint sub-bands. Over $M$ messages, center the stacked $\{V_m\}$ into the
		$M{\times}F$ matrix $D$; split the $F$ bins into $K$ bands and let $U_b$ be the top-$\kappa$ left
		singular vectors of $D$ restricted to band $b$; then
		\begin{equation}
			\mathrm{Coherence}=\mathrm{mean}_{b<b'}\tfrac1\kappa\lVert U_b^{\top}U_{b'}\rVert_F^2,
			\label{eq:coherence}
		\end{equation}
		with chance level $\kappa/M{\approx}0.14$. High $\Rightarrow$ a frequency-redundant payload, so a single notch leaves a
		surviving sub-band and only a \emph{band-sweep} (low-/high-/band-pass) erases it.
		\item \textbf{Estimability}: whether a fixed, host-independent magnitude template transfers across clips. For
		each clip form the unit residual-magnitude footprint $v_i$ and project out the mean host envelope
		direction $\bar H$, $R_i{=}v_i{-}(v_i^{\top}\bar H)\bar H$; then
		\begin{equation}
			\mathrm{Estimability}=\mathrm{mean}_{i<j}\hat R_i^{\top}\hat R_j,
			\label{eq:estimability}
		\end{equation}
		with unit-normalized $\hat R_i$. High $\Rightarrow$
		a universal additive carrier estimable from a few pairs $\Rightarrow$ \emph{template subtraction}.
		\item \textbf{Linearity}: additivity of the embedding in the payload bits. For bit $i$ and context message
		$m^{(c)}$, let $\delta_i^{(c)}{=}\mathrm{embed}(x,m^{(c)}{\oplus}e_i){-}\mathrm{embed}(x,m^{(c)})$ be the
		single-bit-flip delta; then
		\begin{equation}
			\mathrm{Linearity}=\mathrm{mean}_i\,\mathrm{mean}_{c<c'}
			|\cos(\delta_i^{(c)},\delta_i^{(c')})|.
			\label{eq:linearity}
		\end{equation}
		High $\Rightarrow$ a context-invariant linear axis $\Rightarrow$
		an \emph{estimation-subtraction} attack applies (estimate the linear footprint and subtract it);
		low $\Rightarrow$ not linearly attackable.
	\end{itemize}
	Fig.~\ref{fig:signatures} reports the probe values and the signature pattern: each scheme's defining probe
	selects its matched attack, and a clear fragile/robust split emerges (e.g.\ AudioSeal's high
	Concentration/Estimability, DNN-AWM's lone-high Coherence, Timbre's high Linearity, the latent marks' uniformly
	low values), which is what the matched attacks and the scheme identifier exploit.
	
	\section{Probe-Matched Attacks}\label{sec:matched}
	For each watermark the diagnosed embedding domain selects a matched attack. We report the gentlest
	(highest-quality) setting that removes the payload (bit accuracy $<70\%$, chance $\approx 50\%$) and/or the
	detection flag. Table~\ref{tab:matched} summarizes; all attacks here are energy/magnitude-domain, so PESQ is a
	valid quality measure.
	
	\textbf{From probe values to an attack.} We read the four probes as a \emph{joint signature} placing each
	scheme in an archetype with a matched attack (Table~\ref{tab:archetype}), never thresholding one probe alone:
	AlignMark's $0.62$ and AWARE's $0.43$ Estimability sit between the fixed-template cluster ($\ge0.78$) and the
	latent marks ($\le0.28$), yet their payloads resist subtraction. Hence we classify, then verify by running the
	attack, and predicting an \emph{unseen} scheme's class is only partly reliable (Sec.~\ref{sec:limits}).
	
	\begin{table}[t]\begin{center}\begin{threeparttable}
				\caption{Joint probe signature to matched attack. We assign an archetype, then verify by running the attack.}
				\label{tab:archetype}
				\footnotesize\setlength{\tabcolsep}{3pt}\renewcommand{\arraystretch}{1.1}
				\begin{tabular}{|>{\raggedright\arraybackslash}p{0.30\columnwidth}|>{\raggedright\arraybackslash}p{0.22\columnwidth}|>{\raggedright\arraybackslash}p{0.30\columnwidth}|}
					\hline
					Archetype (defining probe) & Matched attack & Schemes \\\hline
					\emph{Narrowband carrier} (Concentration) & band-stop & AudioSeal \\\hline
					\emph{Frequency-redundant} (Coherence) & band-sweep & DNN-AWM \\\hline
					\emph{Universal template} (Estimability) & template subtraction & audiowmark, WavMark, SilentCipher, AudioSeal payload \\\hline
					\emph{Linear in payload bits} (Linearity) & estimation subtraction & Timbre \\\hline
				\end{tabular}
	\end{threeparttable}\end{center}\end{table}
	
	\textbf{WavMark $\rightarrow$ magnitude subtraction.} WavMark embeds in the STFT magnitude; a universal
	magnitude template estimated from pairs and subtracted (over-subtraction) drives the payload to
	$5.9\%$ (near-complete erasure; chance $50\%$) at speech PESQ $4.17$, also dropping detection.
	
	\textbf{SilentCipher $\rightarrow$ content-adaptive subtraction.} SilentCipher scales its magnitude mark by
	per-frame RMS; a content-adaptive (RMS-weighted) subtraction matched to that scaling drives the payload to
	$10.3\%$ at speech PESQ $3.84$, where a fixed template does not.
	
	\textbf{Timbre $\rightarrow$ estimation subtraction.} Timbre is trained robust to signal-level
	distortions, but its embedding is additive and near-linear in the payload bits (high Linearity): the
	watermark's time-averaged magnitude footprint decomposes into a message-independent carrier plus per-bit
	directions, both fit offline from encoder pairs. We estimate the footprint present in the target and subtract it from every frame's magnitude,
	keeping phase. The attack is fully black-box: over
	diverse content it dents the payload to $62.6\%$ (reaching chance on the most favorable clips) at speech
	PESQ $3.99$ (audio-mode ViSQOL $4.30$), where naive full-band optimization
	crushed PESQ to ${\sim}1.8$.
	
	\textbf{AudioSeal $\rightarrow$ notch (detection) and template subtraction (payload).} AudioSeal's residual is
	concentrated in a ${\sim}1.25$\,kHz band (high Concentration; $88\%$ of residual energy in one $1/16$ band); a
	matched band-stop notch there removes the \emph{detection} flag (to $0.05$) transparently, but even swept to
	the transparency limit (speech PESQ $3.62$, ViSQOL $4.55$) it drives the payload only to $60.7\%$ (chance
	$50\%$). Concentration buys the presence flag; the message falls to AudioSeal's \emph{other} high probe,
	Estimability ($0.95$): template subtraction reaches $52.4\%$ -- chance -- at detection $0.01$ and at
	\emph{higher} quality than the notch (speech PESQ $3.89$, ViSQOL $4.63$). Two probes fire and each selects the
	attack that removes a different component, which is the joint signature (Table~\ref{tab:archetype}) working as
	intended.
	
	\textbf{DNN-AWM $\rightarrow$ band-sweep (low-pass).} DNN-AWM has the highest Coherence ($0.62$): its payload
	is replicated across frequency bands, so a notch leaves a surviving sub-band that re-decodes and template
	subtraction barely dents it. The Coherence-matched move is a band-sweep, a low-pass that removes the
	\emph{entire} occupied band at once. A low-pass at \DNNAWMcut\ drives the payload to \DNNAWMacc\ at PESQ
	\DNNAWMpesq\ (speech), where the magnitude attacks failed, lifting DNN-AWM's fragility from $0.45$ under
	subtraction to \DNNAWMfrag. It is the \emph{borderline} case: Coherence surfaces the only training-free attack
	that reaches its payload, but removal lands just under the $>3.5$ transparency line.
	
	\textbf{VoiceMark $\rightarrow$ resists (anchor).} VoiceMark's uniformly low probe values (low Estimability,
	low Linearity, no concentrated or redundant band) predict that no training-free magnitude attack should
	work, and none does: the matched subtraction only dents it to $61.9\%$ by destroying quality (speech PESQ $1.52$). The
	probes that say ``robust'' are correct, which we take as a useful negative.
	
	\textbf{Remaining roster.} The same rule classifies the rest, and the fragility scores
	(Table~\ref{tab:matched}) confirm the predicted split. audiowmark, a grid-locked magnitude scheme, breaks
	cleanly: its matched subtraction drives the payload to $48\%$ and removes detection at speech PESQ $4.22$.
	The remaining marks AlignMark,
	WMCodec, and AWARE resist: the matched subtraction dents their payload only by driving quality down (speech
	PESQ $1.5$--$2.4$), so they join VoiceMark as anchors (low fragility, $0.27$--$0.39$). Table~\ref{tab:matched}
	reports the matched attack and operating point for all ten schemes (break for the fragile group,
	``resists'' for the anchors). Per content type, ViSQOL stays within $4.11$--$4.70$ for every removed scheme,
	so the result is not a speech artifact; the sole exception is DNN-AWM at $2.78$ on music, where its
	band-sweep removes real musical bandwidth.
	
	\begin{table}[t]\begin{center}\begin{threeparttable}
				\caption{Probe-matched removals. Lower bit-acc $=$ payload removed (chance $\approx50\%$); det $=$ detection
					flag there. Means $\pm95\%$ CI, $N{=}100$ clips ($89$ for \texttt{audiowmark}; $84$ for its fragility).
					\textbf{PESQ is speech content only} ($N{=}60$); audio-mode ViSQOL covers all five content types.}
				\label{tab:matched}
\scriptsize\setlength{\tabcolsep}{1pt}
\begin{tabular}{|l|l|c|c|c|c@{\hspace{1pt}}|c@{\hspace{1pt}}|}
\hline
\rule{0pt}{2.2ex}Scheme & matched attack & bit-acc & det & PESQ & ViSQOL & frag\tnote{a} \\\hline
\multicolumn{7}{|l|}{\emph{Removed at near-transparent quality:}}\\\hline
audiowmark           & template subtr.           & $\mathbf{48.4\pm2.9}$ & \textbf{0.00} & 4.22 & 4.62 & $0.85\pm0.03$ \\
WavMark              & template subtr.           & $\mathbf{5.9\pm4.6}$ & \textbf{0.06} & 4.17 & 4.47 & $0.87\pm0.05$ \\
AudioSeal            & band-stop                 & $60.7\pm2.8$ & \textbf{0.05} & 3.62 & 4.55 & $0.80\pm0.03$ \\
SilentCipher         & RMS-adaptive subtr.       & $\mathbf{10.3\pm4.1}$ & 0.20 & 3.84 & 4.22 & $0.84\pm0.03$ \\
DNN-AWM\tnote{b}     & band-sweep                & $58.9\pm3.6$ & 1.00 & 3.34 & 3.36 & $0.62\pm0.04$ \\
Timbre               & estimation subtr.         & $62.6\pm1.2$ & 1.00 & 3.99 & 4.30 & $0.74\pm0.04$ \\
\hline
\multicolumn{7}{|l|}{\emph{Robust anchors (mark only dents at audible quality):}}\\\hline
WMCodec              & template subtr.           & $57.4\pm2.1$ & 1.00 & 1.47 & 3.27 & $0.27\pm0.04$ \\
VoiceMark            & template subtr.           & $61.9\pm2.6$ & 0.54 & 1.52 & 3.50 & $0.28\pm0.05$ \\
AWARE                & template subtr.           & $81.0\pm2.5$ & 0.09 & 2.39 & 4.28 & $0.38\pm0.05$ \\
AlignMark            & template subtr.           & $61.3\pm2.6$ & 1.00 & 1.69 & 3.77 & $0.39\pm0.06$ \\
\hline
\end{tabular}

				\begin{tablenotes}
					\item[a] Threshold-free fragility, (\ref{eq:fragility}); one attack family per scheme.
					\item[b] Borderline: removed, but just under the $>3.5$ transparency line.
				\end{tablenotes}
	\end{threeparttable}\end{center}\end{table}
	
	The matched attacks also expose a \emph{payload/presence} split: for Timbre and AudioSeal the message and the
	detection flag are removed by different means (Timbre's payload falls while detection persists; AudioSeal's
	detection falls to its Concentration-matched notch and its payload to its Estimability-matched subtraction).
	
	\begin{figure}[t]\begin{center}
			\includegraphics[width=85mm]{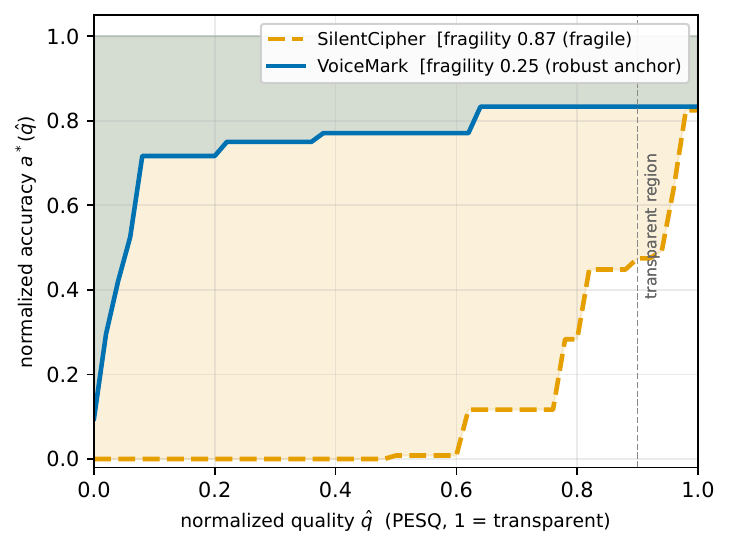}
		\end{center}
		\caption{\textbf{Threshold-free fragility} ((\ref{eq:fragility})): the area \emph{above} the
			accuracy--quality frontier, traced by sweeping the matched attack ($\alpha\in\{1,\dots,16\}$).
			Per-content curves for SilentCipher (fragile, orange dashed) and VoiceMark (robust, blue solid);
			cf.\ the content-averaged $0.84$/$0.28$ in Table~\ref{tab:matched}. A fragile mark's accuracy
			collapses while quality is still high; a robust mark's persists until the audio is destroyed.}
		\label{fig:frontier}
	\end{figure}
	
	\subsection{Quantifying fragility}\label{sec:fragility}
	Reporting accuracy alone is incomplete: a watermark is fragile only if its accuracy collapses \emph{while
		audio quality stays high}. Sweeping the matched attack traces a frontier of (quality, accuracy) points; we
	summarize it with one number and \emph{no} threshold. Let $\mathrm{chance}=50\%$ be the random-guess bit
	accuracy and $\mathrm{clean}$ the watermark's pre-attack accuracy. For each attacked point we define a
	normalized quality $\hat q\in[0,1]$ from any perceptual quality measure $Q$, rescaled to its valid range as
	$\hat q=(Q-Q_{\min})/(Q_{\max}-Q_{\min})$; the choice of $Q$ is use-case dependent; we use PESQ
	($\hat q=(\mathrm{PESQ}-1)/3.5$), but ViSQOL or a subjective listening test (e.g.\ MUSHRA) can also be used while keeping slot unchanged.
	We pair it with a normalized accuracy
	$\hat a=(\mathrm{acc}-\mathrm{chance})/(\mathrm{clean}-\mathrm{chance})\in[0,1]$. The attacker's frontier
	$a^\ast(q)=\min\{\hat a : \hat q\ge q\}$ is the best (lowest) accuracy reachable while holding quality at
	least $q$, and $R=\int_0^1 a^\ast(q)\,dq$ is the area under it. The fragility is then
	\begin{equation}
		\mathrm{fragility} = 1-R \in[0,1],
		\label{eq:fragility}
	\end{equation}
	where $R$ is the frontier area defined above; fragility
	$1$ means the message is erased at near-transparent quality, $0$ means robust.
	Two caveats: fragility is a \emph{lower bound} measured w.r.t.\ the swept
	arsenal, and the numbers we report instantiate $Q$ with PESQ, so they cover only energy/magnitude attacks
	(where PESQ is valid; sync excluded). AudioSeal makes the first caveat concrete. We score it on the
	Concentration-matched notch, which costs \emph{no} fitting queries, but on a like-for-like sweep its
	$15$-query template subtraction reaches $0.87{\pm}0.03$ against the notch's $0.66{\pm}0.04$: the two probes
	buy different things, and the tabulated $0.80$ is a floor set by the cheapest attack rather than AudioSeal's
	best-known removability.
	Fig.~\ref{fig:frontier} illustrates the metric on measured data, contrasting the frontier curves of a fragile
	mark (SilentCipher) and a robust anchor (VoiceMark): the shaded area above each curve is its fragility. The
	figure plots the per-content frontier (fragility $0.87$ for SilentCipher, $0.25$ for VoiceMark), whereas
	Table~\ref{tab:matched} reports the content-averaged value ($0.84$ and $0.28$); the small offset is the
	averaging, and the fragile/robust ordering is identical either way.
	The per-scheme scores in Table~\ref{tab:matched} confirm the split: the magnitude-template and narrowband
	marks (audiowmark, WavMark, AudioSeal, SilentCipher) are fragile ($\to\!1$), DNN-AWM is the borderline
	magnitude mark reachable only by its Coherence-matched band-sweep (fragility \DNNAWMfrag), and the latent-domain
	marks (VoiceMark, WMCodec, AWARE, AlignMark) stay robust ($\to\!0$), a single-number verdict, stable across
	content, that an accuracy-only benchmark cannot give. We read \texttt{audiowmark}'s high score ($0.85$)
	conservatively: it is a detect-or-nothing scheme whose block detector the subtraction desynchronizes, scored
	only over clips where it embeds. Robust latent marks also read anomalously fragile on singing, where weak
	embedding shrinks the $(\text{clean}-\text{chance})$ denominator: an embedding, not a removal, effect, and one
	the content-averaged scores absorb.
	
	\textbf{Identifying the scheme.} Beyond selecting attacks, the same per-clip probe features identify \emph{which}
	scheme is present. A random-forest classifier on the three per-clip probes (Concentration, Coherence, Linearity;
	Estimability is excluded as a per-scheme constant to avoid label leakage) reaches $83.7\%$ accuracy over ten
	schemes (chance $10.0\%$) under group cross-validation by host clip. The systematic confusions
	(audiowmark$\leftrightarrow$WMCodec, AlignMark$\leftrightarrow$VoiceMark) pair low-energy, near-chance-Linearity
	schemes, mechanistically sensible errors that confirm the structure the matched attacks exploit.
	
	\section{Scope and Limitations}\label{sec:limits}
	\textbf{Synchronization excluded: no valid quality metric.} We deliberately exclude synchronization attacks. A tiny, inaudible shift can remove the grid-locked schemes' payload
	(AudioSeal, \texttt{audiowmark}), but we cannot certify it transparent: PESQ and audio-mode ViSQOL both
	time-align internally and over-rate desynchronization, so neither our quality measure nor the PESQ/ViSQOL-based
	fragility score is meaningful there; only a subjective listening test would be. We therefore lead with the
	energy/magnitude breaks, where both metrics are admissible, and no probe maps to the sync axis.
	\textbf{Training-free, encoder-only attacks.} Every probe and attack is built from clean/watermarked pairs alone: closed-form moves (magnitude subtraction, filtering, footprint
	estimation-subtraction), never gradients, decoder access, or a trained removal network. Each scheme is removed by a \emph{single} matched attack family, so the fragility scores are a
	\emph{conservative} lower bound on removability: the concurrent HarmonicAttack~\cite{harmonicattack} trains a
	single \emph{universal} removal network on pairs and reports $100\%$ attack success against all four schemes it
	evaluates (AudioSeal, WavMark, SilentCipher, AudioMarkNet) on unseen music at ViSQOL ${\approx}4.33$, so even
	our probe-robust anchors may fall to a learned remover. The cost, however, differs in kind: their released implementation is a
	$38.0$\,M-parameter dual-path autoencoder trained GAN-style against a $0.8$\,M-parameter discriminator, and at
	our $4$\,s, $16$\,kHz operating point that network costs about $233$\,GFLOP per clip. Ours trains nothing, holds
	\emph{no} parameters, and costs $17$\,MFLOP per clip -- a factor of $1.4{\times}10^{4}$ -- after a one-off fit
	of at most $150$ encoder queries. Our contribution is
	orthogonal: a \emph{diagnostic} account of \emph{why} each mark falls and to \emph{which} perturbation, at
	near-zero cost.
	
	\subsection{Implications for watermark design}\label{sec:mitigate}
	Read in reverse, each probe names a property to avoid and implies its remedy.
	\emph{Concentration}: spread the mark across the spectrum and make the detector require wideband evidence.
	\emph{Coherence}: bind the payload jointly across bands so that no sub-band decodes alone --- per-band
	replication buys less than it appears to, since on DNN-AWM at matched quality a notch leaves the payload
	untouched ($100\%$) where a sweep reaches $74.5\%$. \emph{Estimability}: make the mark host-adaptive and
	content-keyed, so that no universal template exists. \emph{Linearity}: embed bits non-linearly and
	context-dependently, so that per-bit directions cannot be fitted offline. We do not build and test a hardened
	scheme, so these are implications rather than validated defences.
	Empirically every removed scheme here has a probe at $0.67$ or above and no robust scheme exceeds $0.62$ --- an
	observed separation over ten schemes, not a threshold. The probes are cheap enough to serve as a
	\emph{pre-deployment self-audit} of one's own encoder before release.
	
	\section{Conclusion}
	This work frames watermark removal as a diagnostic problem: probing the embedding domain from
	clean/watermarked pairs and applying a matched attack accordingly. Across ten watermarking schemes (spanning
	magnitude-, vocoder-, codec-latent-, and speaker-domain marks, plus the classical tool
	\texttt{audiowmark}), the framework removes the fragile schemes at high objective quality (PESQ/ViSQOL) and
	exposes the latent-domain marks as robust for non-sync training free attacks, both outcomes consistent with what the probes indicate and
	informative about the embedding strategy.
	Future work will extend the framework to additional schemes, add perceptual evaluation of
	synchronization-based attacks, and explore watermark forgery (overwriting) under the same diagnostic framework.

	\section*{Acknowledgment}
	The author used an AI assistant (Claude) for software implementation of the experimental pipeline, manuscript editing, and LaTeX formatting. All research design, experimental methodology, results, and conclusions are the author's own; the author verified all reported results and takes full responsibility for the content.
	
	\balance
	\printbibliography
	
\end{document}